# Nonlinear second-order photonic topological insulators

Marco S. Kirsch[1,†], Yiqi Zhang[2,†], Mark Kremer[1], Lukas J. Maczewsky[1], Sergey K. Ivanov[3,4], Yaroslav V. Kartashov[3,5], Lluis Torner[5], Dieter Bauer[1], Alexander Szameit[1,*], and Matthias Heinrich[1]

[1]Institut für Physik, Universität Rostock, Albert-Einstein-Str. 23, 18059 Rostock, Germany

[2] Key Laboratory for Physical Electronics and Devices of the Ministry of Education & Shaanxi Key Lab of Information Photonic Technique, School of Electronic Science and Engineering, Xi'an Jiaotong University, Xi'an 710049, China

[3]Institute of Spectroscopy, Russian Academy of Sciences, Fizicheskaya Str., 5, Troitsk, Moscow, 108840, Russia

[4]Moscow Institute of Physics and Technology, Institutsky lane 9, Dolgoprudny, Moscow region, 141700, Russia

[5]ICFO-Institut de Ciencies Fotoniques, The Barcelona Institute of Science and Technology, 08860 Castelldefels (Barcelona), Spain

[†]These authors contributed equally to this work

[*]Correspondence to: alexander.szameit@uni-rostock.de

## Abstract

**Higher-order topological insulators (HOTI) are a novel topological phase beyond the framework of the conventional bulk-boundary correspondence [1,2]. In these peculiar systems, the topologically nontrivial boundary modes are characterized by a co-dimension of at least two [3,4]. Despite several promising preliminary considerations regarding the impact of nonlinearity in such systems [5,6], the flourishing field of experimental HOTI research has thus far been confined to the linear evolution of topological states. As such, the observation of the interplay between nonlinearity and the dynamics of higher-order topological phases in conservative systems remains elusive. In our work, we experimentally demonstrate nonlinear higher-order topological corner states. Our photonic platform enables us to observe nonlinear topological corner states as well as the formation of solitons in such topological structures. Our work paves the way towards the exploration of topological properties of matter in the nonlinear regime, and may herald a new class of compact devices that harnesses the intriguing features of topology in an on-demand fashion.**

Topological insulators are a recently discovered state of matter. Among their unique features are chiral surface currents that are topologically protected from scattering at defects and disorder, while the bulk material remains insulating [7]. Soon after the first experimental realizations in condensed matter systems, topological notions proliferated across other fields of physics, resulting in the experimental demonstrations of topological dynamics in various platforms, particularly in photonics [8,9]. Recently, as a generalization to these concepts, the existence of higher-order topological insulators (HOTI) was proposed [1,10] and experimentally demonstrated in solid state systems [11] as well as a plethora of other platforms, including phononics [12] and photonics [13-15]. In this new type of topological phase, the dimensionality of the topologically nontrivial boundary modes is more than one dimension below that of the bulk. In other words, a $d$-dimensional $n$-th order topological insulator supports $(d-n)$-dimensional boundary states [4] (see Fig. 1a). Although the study of HOTI is still in its infancy, the applicative potential of these systems has been widely recognized, e.g. for robust high-Q cavities [16] that harness the unusually strong modal confinement of corner states. Moreover, HOTIs are also related to real-space topological defects such as lattice disclinations [17], which may enable the realization of Majorana bound states and non-Abelian braiding statistics [18].

Despite a number of promising preliminary theoretical considerations [6,19,20], several experimental studies in dissipative HOTI systems [21-23], and the recently demonstrated potential of harnessing tightly localized HOTI corner states for the enhancement of nonlinear processes such as third-harmonic generation [24], there is only one experimental paper at hand, where evidence of switching into the HOTI phase under the nonlinear action of a homogeneous global pump was reported in electronic circuits [5]. Importantly, to date, research into the wave dynamics of HOTIs in conservative optical systems has been confined to strictly linear conditions, and their extension in the presence of nonlinear self-action has yet to be explored. Yet, even these first glances into the nonlinear regime raise various intriguing fundamental questions. In particular, as the concept of band topology has thus far been considered to be inherently tied to linear systems, the generalization to nonlinear systems is by no means straightforward. Since topological properties can indeed persist in the presence of nonlinearity [25] or even be established by it [26], does the same hold true for HOTI?

In this work, we experimentally explore the nonlinear dynamics of light in photonic HOTIs based on the Kagome geometry (see Fig. 1b). The topological phase of this lattice type is characterized by the corresponding topological invariants - two bulk polarizations $\mathcal{P}_x = \mathcal{P}_y = \frac{1}{3}$ [27,6,15]. We observe the emergence of nonlinear topological corner states as they bifurcate from their linear counterpart, as well as the formation of spatial solitons in such structures. To this end, we employ lattices of evanescently coupled optical waveguides, characterized by a focusing Kerr response, as versatile testbed for nonlinear physics [28,25,26].

The wave dynamics in our system obey the normalized continuous nonlinear Schrödinger equation for the dimensionless light field amplitude $q$ propagating along the normalized longitudinal coordinate $\zeta$:

$$i\frac{\partial q}{\partial \zeta} = -\left(\frac{\partial^2}{\partial \xi^2} + \frac{\partial^2}{\partial \eta^2}\right)q - |q|^2 q - pR(\xi,\eta)q . \qquad (1)$$

Here, $\xi$ and $\eta$ are the normalized transverse coordinates, the parameter $p$ describes the contrast of the refractive index modulation, whereas the lattice profile itself is characterized by the function $R(\xi,\eta)$. The expression $|q|^2 q$ describes the intensity-dependent shift of the local propagation constant as mediated by the action of the Kerr-type nonlinearity [28]. For low input powers, this term is negligible, and the light evolution is effectively linear. For increasing input power, however, the intensity-dependent phases may lead to significant substantial changes of the propagation dynamics across the lattice sites. We note that Eq. (1) is formally equivalent to the well-known Gross-Pitaevskii equation, which describes the mean-field evolution of bosonic many-particle quantum systems with inter-particle interaction. In this vein, our classical photonic system also offers a model capable of probing the evolution of many-body quantum states.

The topological characteristics of the Kagome lattice, arising from the degree of dimerization of the bond strengths within and between the individual unit cells, are summarized in Fig. 1b. We here consider a semi-infinite Kagome lattice truncated along two intersecting lattice planes such that a 60° corner is formed on its upper left. In Fig. 1b, the limiting cases of dimerization are illustrated. When the intra-cell bond $C_1$ is weaker than the inter-cell bond $C_2$ (left panel), the topological phase of the system is associated with emerging non-trivial polarizations $\mathcal{P}_x = \mathcal{P}_y = \frac{1}{3}$ [22,6,15]. The trivial phase, identified by vanishing polarizations $\mathcal{P}_x = \mathcal{P}_y = 0$ [22,6,15], occurs when the strength of the intra-cell bond exceeds that of the inter-cell bond ($C_1 > C_2$, right panel). The central panel shows the ensemble of propagation constants $\beta$ (representing the tight-binding eigenvalues of the linear version of Eq. (1) for a triangular Kagome plaquette comprised of 84 sites) as a function of the dimerization. The homogeneous lattice is defined by a value of 0.5 for the dimerization parameter $\Delta = C_1/(C_1 + C_2)$, marking the phase transition between the topological regime ($\Delta < 0.5$) and the trivial one ($\Delta > 0.5$). Exemplary mode fields corresponding to the numbered locations in this "band structure" are shown in Fig. 1c. Indeed, only the topological regime allows for states whose eigenvalues fall in the topological gap separating the bulk bands of the lattice for $\Delta < 0.5$. Whereas the branches associated with the conventional topological edge states exhibits a certain slope, the mid-gap branch is exclusively composed of states residing at the corners of the lattice. It is exactly these corner states that arise due to the higher-order topology of the lattice, and, as such, constitute the defining feature of our second-order photonic topological insulator: In contrast to conventional TIs, such as insulators with broken time-reversal symmetry supporting chiral currents at their edges, HOTIs may support not only edge or hinge states, but also immobile topological corner modes (Fig. 1a).

In order to experimentally probe the linear as well as nonlinear dynamics, we fabricated rhomboidal Kagome plaquettes comprised of 65 waveguides with the femtosecond laser direct writing technique, and excited the corner sites with intense ultrashort laser pulses to provide sufficient peak powers to elicit a nonlinear response. Figure 2 presents an overview of characteristic output patterns after a sample length of 100mm for different dimerization as well as different input powers. In all cases, light was injected into the upper left corner site. To ensure a common power scale despite the widely varying absolute coupling values present for the samples with different dimerizations, the injected powers $P_{\text{in}}$ are normalized with respect to the sum of the two couplings, i.e. $P = P_{\text{in}}/(C_1 + C_2)$. The experimentally observed results are shown in Fig. 2a, whereas Fig. 2b provides the corresponding numerical simulations, where the numerically launched dimensionless power $U_{\text{in}}$ is equivalently normalized as $U = U_{\text{in}}/(C_1 + C_2)$. In the homogeneous lattice ($\Delta = 0.5$), a linear excitation ($P = 0.2 \text{MW} \cdot \text{cm}$) penetrates deeply into the lattice. When increasing the power to $P = 1.6 \text{MW} \cdot \text{cm}$, a decrease in transverse broadening marks the onset of nonlinearity, and at even higher high input powers ($P = 2.5 \text{MW} \cdot \text{cm}$), light remains confined to the initially excited waveguide as a tightly confined soliton has formed [29]. For increasing dimerization $\Delta > 0.5$, this behavior remains qualitatively similar, as the lattice continues to reside in the topologically trivial phase with $\mathcal{P}_x = \mathcal{P}_y = 0$: Linear excitations diffract widely across the lattice, and localization gradually increases with the injected power as the output distribution contracts towards the excited corner, even if the complete collapse onto a single waveguide shifts towards the higher powers necessary to overcome the larger intra-cell coupling $C_2$. This behavior changes significantly for dimerizations $\Delta < 0.5$, where the lattice is in the topological phase and $\mathcal{P}_x = \mathcal{P}_y = \frac{1}{3}$. Deeply in the topological phase (at $\Delta = 0.2$), the tightly confined topological corner state captures the vast majority of the launched light, and the bulk of the lattice remains essentially dark. Yet, as the input power is increased, the weakly-nonlinear regime ($P = 1.5\text{MW} \cdot \text{cm}$) actually yields an intermediate delocalization, where nonlinear phase matching to the topological edge states as well as the bulk bands allows a certain fraction of light to escape from the corner. At even higher powers, the strongly nonlinear regime ($P = 4.0\text{MW} \cdot \text{cm}$) finally all serves to trap light in the outermost corner site of the plaquette.

Figure 3 depicts the power-dependent localization behavior in greater resolution. The relative fraction of the overall intensity remaining in the corner waveguide serves as quantitative measure for the suppression of broadening of the single-site excitation, and converges to unity for perfect single-site localization. We start our analysis for corner excitations in the lattices with $\Delta > 0.5$ (Fig. 3, right column). The broad linear diffraction patterns at low intensities involve a large portion of the plaquette. Increasing input powers generally lead to increasing localization as the light contracts towards the excited corner site. This is the typical behavior for two-dimensional corner lattice solitons [29]. Notably, comparing the dimerized cases to the homogeneous lattice, one finds that the onset of localization systematically occurs at lower powers for larger $\Delta$, while at the same time higher powers are required completely trap light in the corner site, in line with the two characteristic thresholds defined by the powers necessary to overcome the inter- and intra-cell couplings, respectively.

When launching light into a bulk site (Fig. 3, central column), only the deviation from the homogeneous case ($\Delta = 0.5$) matters, as the distinction between inter- and intra-cell couplings is moot in the absence of boundaries. For weakly dimerized lattices ($0.45 < \Delta < 0.55$), light contracts smoothly with increasing power, as the two couplings in the system are of similar magnitude. As a result, the strongly nonlinear regime with its spatial solitons [30] is quickly reached. Instead, strongly dimerized lattices ($\Delta = 0.2$ and $\Delta = 0.8$) once more show an accelerated onset of nonlinear behavior, while substantially higher powers are required to achieve full localization, as can be seen with particular clarity in the numerical simulations (lower row of Fig. 3).

This picture changes drastically when exciting the corner waveguide in the strongly topological regime ($\Delta = 0.2$, Fig. 3, left column). In this case, light is indeed strongly localized in the linear regime, as the topological corner states is excited – the hallmark of our second-order topological insulator. When increasing the input power, the localization gradually decreases as the weakly-nonlinear corner excitation traverses the window of phase matching with edge and bulk states between $1.0 \text{MW·cm} < P < 2.5 \text{MW·cm}$. Only in the strongly nonlinear regime, when the power ($P > 2.5 \text{MW·cm}$) is sufficient to drive the excitation into the semi-infinite gap above the bands of the lattice, a corner soliton forms and light once more localizes in the corner site.

In addition to a phenomenological description of the dynamics, it is important to note that the corner state indeed remains topological – despite the clear imprint of nonlinearity – until the excitation actually enters the band. In other words, for a sizeable range of powers beyond the linear regime ($0 < P < 1.0 \text{MW} \cdot \text{cm}$), our system remains a second-order topological insulator, and, as such, continues to support a (nonlinear) topological corner state. This can be framed more formally by the following line of thought. For small dimerizations (e.g. $\Delta = 0.2$), the corner state is virtually confined to the outermost unit cell of the topological corner. Therefore, light dynamics in its vicinity are dictated by the local band structure [31,26] that arises from the internal makeup this unit cell. For increasing input powers, self-phase modulation mediated by the Kerr nonlinearity systematically increases the effective refractive index of the excited waveguide. Following the notion that only the local band structure (determined by only that part of the system in which light actually resides) determines the light evolution [31,26], one is at liberty to replace the rest of the (unilluminated) lattice with copies of the effectively detuned unit cell. Under the conditions described above, this substitution facilitates the computation of the effective values for the polarizations $\mathcal{P}_x$, $\mathcal{P}_y$ despite the fact the system to be modelled is actually nonlinear. In this vein, we find that the polarizations are no longer pinned to the value of $1/3$ associated with identical lattice sites, since chiral symmetry is broken by the detuning. Nevertheless, both polarizations remain strictly positive for dimerizations $\Delta < 0.5$ (see Fig. 4). The fact that $\mathcal{P}_x \neq 0$, $\mathcal{P}_y \neq 0$ indicates that the existence of the corner states continues to be a result of the underlying bulk topology even in the presence of high input powers.

In summary, we demonstrated for the first time a nonlinear photonic second-order topological insulator in a conservative physical system. We systematically explored the features of the underlying lattice in the topological phase, and observed soliton formation as well as nonlinear topological corner states in this structure. Based on an integrated-optical platform, these findings open a new experimental avenue towards developing a more holistic understanding of topological insulators and bringing them to application in future compact devices. Indeed, many fascinating directions of inquiry come to mind: In which ways does the introduction of nonlinearity impact exhibiting fermionic time-reversal symmetry [32]? How can gain, loss, and non-Hermiticity in general, be efficiently harnessed to tailor nonlinear wave packet evolution [33]? Does the presence of nonlinearity in non-trivial topologies mitigate or in fact enhance disorder-induced localization mechanisms [34]? The tools to experimentally tackle these questions are now within reach, and will drive the exploration of the scientific as well as technological aspects of nonlinear topology in all kinds of wave mechanical systems, whether in the photonic, acoustic, optomechanical, polaritonic, atomic, or even entirely new domains.

## Acknowledgements

The authors would like to thank C. Otto for preparing the high-quality fused silica samples used for the inscription of all photonic structures employed in this work. The authors acknowledge funding from the Deutsche Forschungsgemeinschaft (grants SCHE 612/6-1, SZ 276/12-1, BL 574/13-1, SZ 276/15-1, SZ 276/20-1), the Alfried Krupp von Bohlen and Halbach foundation, and RFBR (grant 18-502-12080). Y.V.K. and L.T. acknowledge support from the Government of Spain (Severo Ochoa CEX2019-000910-S), Fundació Cellex, Fundació Mir-Puig, Generalitat de Catalunya (CERCA).

## Authors Contributions

MKi, MH and LJM designed and fabricated the photonic lattices. MKi and MH conducted the experiments. YVK, YZ, and SKI were responsible for the continuous simulations. MKi carried out the tight-binding simulations. The experimental and numerical data was evaluated by MKi, MH and LJM. DB and MKr devised the method for the topological characteristics in the nonlinear regime and calculated the polarizations. AS and LT supervised the efforts of their respective groups. All authors co-wrote the manuscript.

## Competing Interests Declaration

The authors declare no competing interests.

## Data and Materials Availability Statement

Further/raw data and numerical code is available on reasonable request.

# Methods

**Waveguide inscription and lattice parameters**

The dimerized Kagome lattices employed in our experiments were fabricated by the femtosecond laser direct writing technique in 100mm long fused silica samples (CORNING 7980). Pulses from a Titanium:Sapphire amplifier system (COHERENT Vitara S/RegA 9000, carrier wavelength 800nm, pulse duration 140 fs, pulse energy 400 nJ, repetition rate 100 kHz) were focused through a microscope objective (20 ×, NA = 0.35). The sample itself was translated relative to the focus with a high-precision positioning system (AEROTECH ALS180) at 100 mm min$^{-1}$, yielding single-mode waveguides (effective refractive index contrast $\approx 5 \times 10^{-4}$, approximate mode fields dimensions 12 μm × 20 μm) for the probe wavelength of 800nm. Waveguide separations between 22.0 μm and 35.7 μm, corresponding to coupling coefficients between 0.20 cm$^{-1}$ and 0.81 cm$^{-1}$, were chosen to implement arrangements with the desired representative dimerizations of $\Delta = 0.2, 0.45, 0.5, 0.55$ and $0.8$.

**Sample characterization**

A high-power Titanium:Sapphire CPA System (COHERENT ASTRELLA-1K-F) provided intense 210fs pulses of up to peak powers of 30GW at the probe wavelength of 800nm and a repetition rate of 1kHz. A microscope objective (2.5 ×, NA = 0.075) was used to inject these pulses into specific lattice sites, and the resulting power-dependent output intensity distributions at the sample end face were imaged onto a CCD camera (BASLER acA1920-155um) by another microscope objective (4 ×, NA = 0.1).

**Calculation of the polarizations**

The topological characterization of the Kagome-type lattices is based on the polarizations $\mathcal{P}_j$, where $j \epsilon \{x, y\}$, which can be numerically calculated by evaluating the integral

$$\mathcal{P}_j = -\frac{1}{4\pi^2} \iint_{BZ} A_j(k) \, dk_x \, dk_y \quad \text{with} \quad A_j(k) = -i \langle u | \frac{\partial}{\partial k_j} | u \rangle,$$

over the Brillouin Zone (BZ), with the Bloch states $|u\rangle$ of the first band [6].

In order to account for the nonlinearity, a detuning $\delta$ is introduced at the site of the unit cell that is excited with a high-power input beam. This approximation is motivated by the consideration that the local band structure describing the dynamics of light [31,26] is in fact determined by only that part of the system in which light actually resides). In this vein, the unilluminated part of the lattice can be substituted with copies of the effectively detuned unit cell for numerical purposes. The corresponding Hamiltonian of this equivalent system then reads

$$H(k) = \begin{pmatrix} 0 & v + w\, e^{i(k_x - k_y)} & v + w\, e^{-ik_y} \\ v + w\, e^{-i(k_x - k_y)} & \delta & v + w\, e^{-ik_x} \\ v + w\, e^{ik_y} & v + w\, e^{ik_x} & 0 \end{pmatrix},$$

where the wave numbers $k_{x/y}$ are normalized to the lattice constant and chosen along orthogonal directions. For the simulations, the resolution of the $k_{x/y}$ along the derivative is chosen to be 3900 points, and 1300 points along the orthogonal direction.

# Figures:

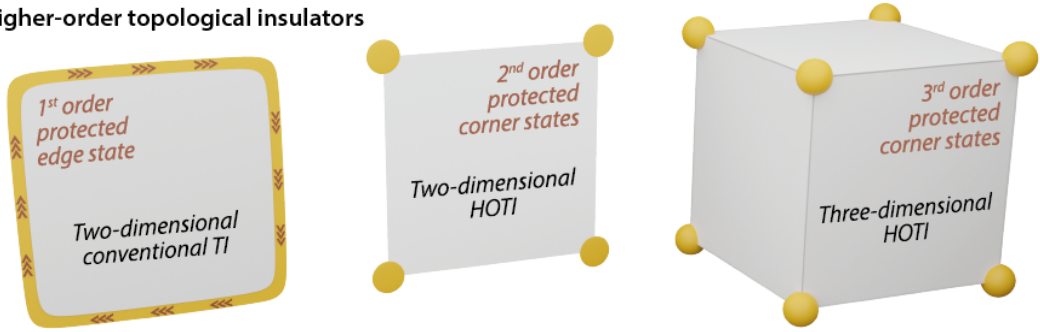

(a) Higher-order topological insulators

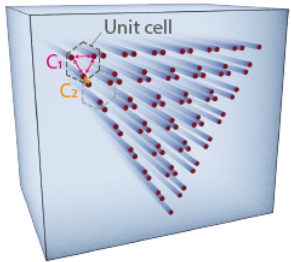
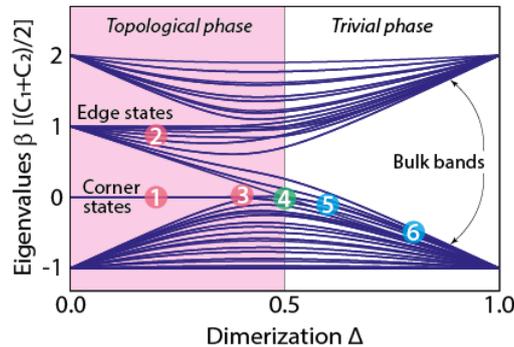
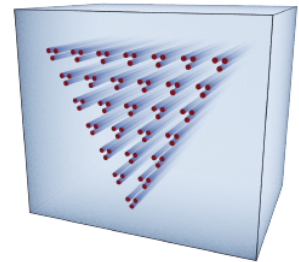

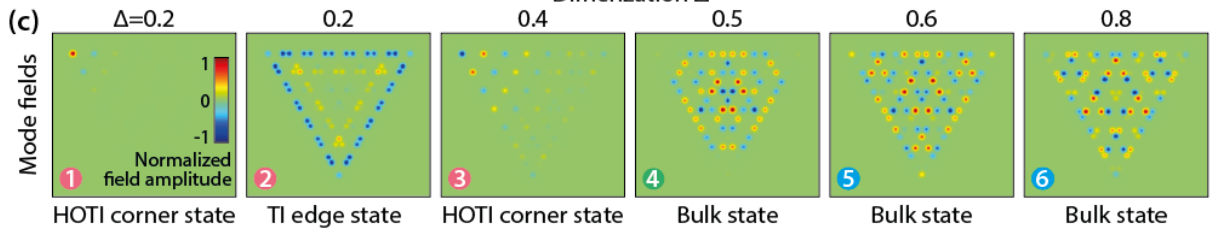

**Figure 1:** (a) Higher-order topological insulators are topological systems in which the protected states in lower-dimensional surface features than in conventional TIs. (b) The Kagome lattice, here shown in its incarnation as two-dimensional array of waveguides, exhibits a topological phase when the coupling $C_2$ between its unit cells exceeds the coupling $C_1$ within them, i.e. for a dimerization $\Delta < 0.5$ (left panel). Plotting the normalized tight-binding eigenvalues of a triangular Kagome plaquette as function of the dimerization (central panel), the branches associated with both topological edge- and HOTI corner-states can be identified. In the trivial regime ($\Delta \geq 0.5$), only bulk bands exist. Numerically computed exemplary mode field distributions corresponding to the marked points within the band plot are shown in (c). As aid to the eye, they have been normalized with respect to their individual maximum absolute values.

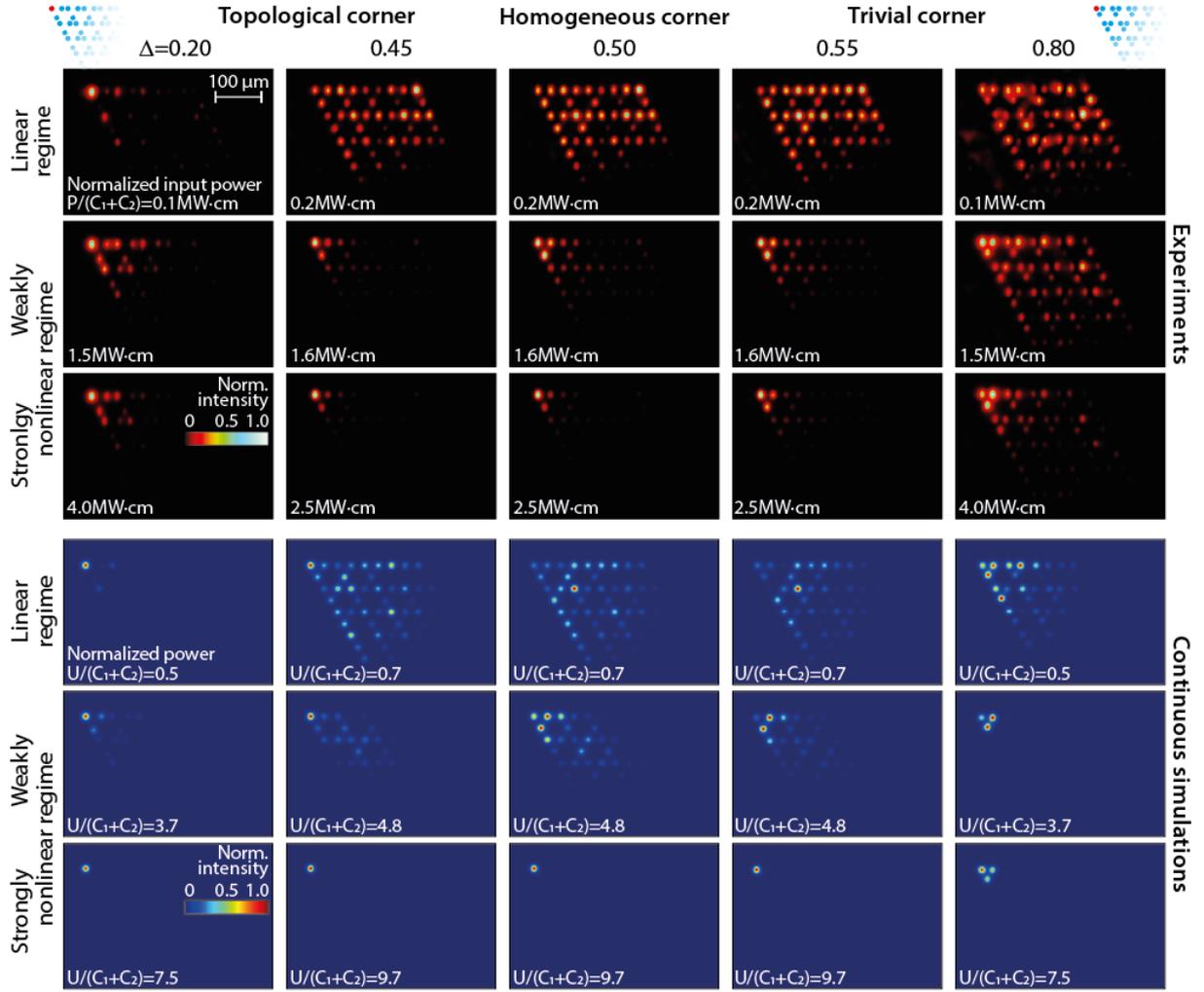

**Figure 2:** Nonlinear corner excitations in dimerized Kagome lattices. Experimental images are shown in the top half, whereas corresponding continuous simulations are shown below. Given the scaling of the nonlinear response with the overall coupling strength in the system, experimental as well as numerical powers are provided in normalized form. First column ($\Delta = 0.2$): In the strongly topological regime, linear corner excitations predominantly populate the HOTI state. Increasing the power leads to an intermediate nonlinear phase-matching to bulk modes, until at high powers, a tightly localized corner soliton forms. Second column ($\Delta = 0.45$): Since the weakly topological regime features HOTI states that extend far into the lattice (cf. Panel 3 in Fig. 1(c)), linear corner excitations yield considerable bulk diffraction, and the additional action of nonlinearity is required to form a localized corner state. Nevertheless, this contraction occurs faster than in the trivial case (middle column, $\Delta = 0.5$). The weakly dimerized trivial regime (fourth column, $\Delta = 0.55$) is instead characterized by intra-unit cell dynamics that yield two separate localization steps as the power is increased. In the strongly dimerized trivial regime (fourth column, $\Delta = 0.8$), the intermediate localization in the three guides of the corner unit cell is even more pronounced.

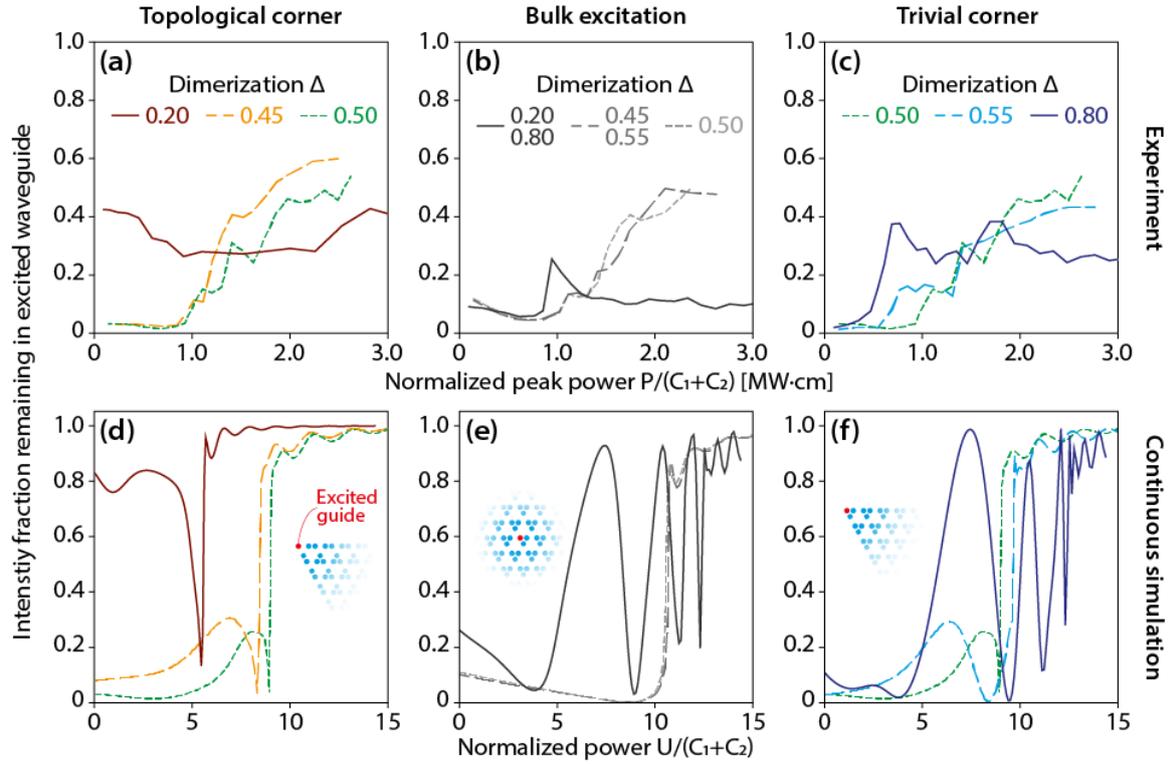

**Figure 3:** Power-dependent localization in the dimerized Kagome lattice. (a) Measured intensity fraction remaining in the topological corner waveguide. In the strongly dimerized case ($\Delta = 0.2$, solid line), nonlinear phase matching to bulk modes temporarily lifts the topological protection of the corner mode, as evidenced by the intermediate delocalization. While the weakly dimerized topological system ($\Delta = 0.45$, long dashed line) does not capture linear single-site excitations at its topological corner, nonlinear localization is accelerated compared to the homogeneous lattice ($\Delta = 0.5$, short dashed line). (b) Bulk excitations cannot distinguish between the topological and trivial regimes. However, the presence of two distinct coupling values in dimerized systems yields two separate power scales at which partial and full localization occurs, respectively. In terms of the normalized power, stronger dimerizations lead to an earlier onset of nonlinear dynamics, but require higher powers to fully localize. (c) A similar two-tiered localization behaviour occurs at trivial corners. (d-f): Corresponding localization plots obtained from continuous numerical simulations. Note that while the linear background inherent in the pulsed excitation scheme employed inevitably smoothes over miscroscopic details in the experimentally observed localization dynamics, the signatures of localization/delocalization thresholds are faithfully reproduced.

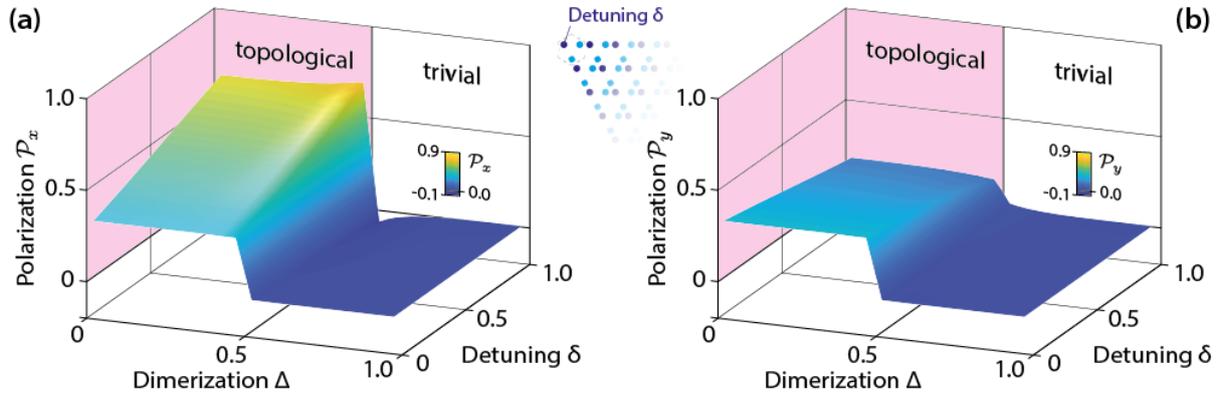

**Figure 4:** The persistence of topological characteristics in the presence of nonlinear dynamics can be probed by considering a Kagome lattice in which one of the three sites of the unit cell features a higher effective index as characterized by the detuning $\delta$. Shown here are the numerically computed polarizations $\mathcal{P}_x$ and $\mathcal{P}_y$ that indicate non-trivial topology when assuming positive values. Starting from the characteristic value of $\frac{1}{3}$ in the homogeneous lattice, both $\mathcal{P}_x$ and $\mathcal{P}_y$ experience a characteristic jump at $\Delta = 0.5$, which marks the emergence of a topological corner state even for large detunings ($\delta = 1$). In other words, the continued existence of the corner states in the nonlinear regime continues to be a result of the underlying bulk topology.